\documentclass[prd, aps,a4paper,showkeys]{revtex4}

\usepackage{amsmath}

\begin{document}

\newcommand\beq{\begin{equation}}
\newcommand\eeq{\end{equation}}
\newcommand\bea{\begin{eqnarray}}
\newcommand\eea{\end{eqnarray}}
\newcommand\bseq{\begin{subequations}} 
\newcommand\eseq{\end{subequations}}

\title{New Issues in the Inflationary Scenario}

\author{Giovanni Imponente}
\affiliation{
School of Mathematical Sciences\\
Queen Mary, University of London\\
Mile End Road, London E1 4NS, UK}
\email{g.imponente@qmul.ac.uk}


\begin{abstract}

We briefly review the arising of an inflationary 
phase in the Universe evolution in order to 
discuss an inhomogeneous cosmological solution 
in presence of a real self 
interacting scalar field minimally coupled 
to gravity in the region of a slow rolling 
potential plateau. 
During the inhomogeneous de Sitter 
phase the scalar field dominant term is 
a function of the
spatial coordinates only. We apply this
generic solution to the Coleman-Weinberg potential 
and to the Lemaitre-Tolman metric.
This framework  specialized nearby the 
FLRW model allows a classical 
origin for the inhomogeneous 
perturbations spectrum.

\end{abstract}
\maketitle


\section{General Statements}

The homogeneous and isotropic Universe so far observed
at sufficient large scales (of the order of 100 Mpc)
inevitably breaks down when considering galaxies, clusters, etc.
The early Universe, as testified by the extreme uniformity 
of the Cosmic Microwave Background Radiation (CMBR), 
exhibits a level of isotropy and homogeneity of order 
of $10^{-4}$; this is  well tested up to $10^{-3} -10^{-2}$ seconds
of its life by the very good agreement between the abundances 
of light elements as predicted by the Standard 
Cosmological Model (SCM)
and the one observed \cite{KT90}.
 
The theoretical framework of the Einstein equations applied 
to cosmology when following the evolution 
towards the initial 
singularity shows instability for density 
perturbations \cite{LK63}, eventually giving 
space to cosmological models differing from the SCM.
Furthermore, reliable indications support the idea that
the Universe evolved through an 
inflationary scenario and since that age it 
reached isotropy and homogeneity  on the 
horizon scale at least \cite{dB01-map}. \\
Despite of these evidences favourable to the large scale 
homogeneity, many relevant 
features suggest that in the very early stage of evolution 
it had to be described by much more general 
inhomogeneous solutions of the Einstein equations.
With respect to this we remark the following two issues:
\begin{enumerate}
\renewcommand{\labelenumi}{({\it\roman{enumi}})}

	\item 	the backward instability of the density perturbations
			to an isotropic and homogeneous Universe \cite{LK63}
			allows us to infer that,
			when approaching the initial singularity, the dynamics 
			evolved as more general and complex models;
			thus the ap\-pea\-rance of an oscillatory regime
			is expected in view of its general nature, 
			i.e.
			because its perturbations correspond simply to 
			redefine the spatial gradients involved in the 
			Cauchy problem.

	\item		Since the Universe dy\-na\-mics during the Planckian 
			era underwent a quantum regi\-me, 
			then no symmetry restrictions can be \textit{a priori}
			imposed on the cosmological model; 
			in fact, the wave functional of the Universe 
			can provide information on the casual scale at most, 
			and therefore the requirement for a global 
			symmetry to hold would imply a large scale 
			correlation of different horizons.
			Thus, the quantum evolution of the Universe is 
			appropriately described only in terms of a generic 
			inhomogeneous cosmological model.

	\end{enumerate}

A peculiar feature of the inflationary scenario
consists of the violent expansion the Universe
underwent du\-ring the de Sitter 
phase \cite{G81, infla}-\cite{BST83};
indeed via such a mechanism the inflationa\-ry 
model provides a satisfactory explanation of 
the so-called \emph{horizons} and \emph{flatness} 
paradoxes by stretching
the inhomogeneities at a very 
large scale \cite{MB95,dB01-map}.
When referred to a (homogeneous 
and isotropic) 
Friedmann--Lemaitre--Robertson--Walker 
(FLRW) model \cite{KT90}, 
the de Sitter phase of the inflatio\-na\-ry 
scenario rules out the small
 inhomogeneous perturbations so strongly, 
 that it makes 
 them unable to become seeds
for the later structures formation \cite{DT94,IM03}.
This picture emerges sharply within 
the inflationary paradigm and it is at 
the ground level of the statements 
according to which 
the cosmological perturbations arise from the 
scalar field quantum fluctuations \cite{STA96}.\\
Though this argument is well settled down 
and results very attractive
even because the predicted quantum spectrum of 
inhomogeneities takes the Harrison-Zeldovich 
form, nevertheless the question 
whether  it is possible, in a more general context,
that classical inhomogeneities survive 
up to a level relevant for the origin of the actual 
Universe large scale structures remains open, 
as recently presented \cite{IMmpla04}.\\

Indeed here we show  the behaviour 
of an inhomogeneous cosmological
model \cite{MAC79-BKL82}
which undergoes a de Sitter phase \cite{SKB}
and show
how such a general scheme allows the scalar field to 
retain, at the end of the exponential expansion, 
a generic inhomogeneous term to leading order 
(for connected topic see \cite{SKB}).\\
Thus our analysis provides relevant information either 
with respect to the morpho\-lo\-gy of an inhomogeneous  
inflationary model, either stating that 
the scalar field
is characterized by an arbitrary spatial function 
which plays the role of its leading order.\\
The paper is structured as follows: 
in Section II we  review 
the basic equations of the SCM and the arising 
of the typical 
inflationary behaviour of the metric scale factor. 
In Section III we introduce the formalism of \cite{BST83},
in terms of the invariant function $\zeta$, to treat  
density perturbations in an inflationary scenario. 
In Section IV we will show the dynamics of 
an inhomogeneous cosmological model coupled with a 
real self-interacting 
scalar field. The solution concerns 
the phase when  
the scalar field slow rolls on a potential  plateau 
and the Universe evolution is dominated by 
the effective cosmological constant related to 
the energy level over the true vacu\-um state of the theory.
In Section V.A we specify such a framework to the Coleman-Weinberg
expression for the potential, while in Section V.B we consider
also the explicit spherically symmetric Tolman-Bondi metric. 
Then we apply in Section V.C the density perturbations estimate
to the FLRW Universe and in Section VI we draw some conclusions.\\
In the overall work, we can neglect the contribution of the 
ultra-relativistic matter which is relevant only 
for higher order terms and becomes more and more
negligible as the exponential expansion develops
(for a discussion of an inflatio\-na\-ry scenario with 
relevant ultra-relativistic matter and different
outcoming behaviour, see \cite{DT94,IM03}).

\section{From the SCM to Inflation}

\subsection{The Friedmann Equation}

The FLRW me\-tric is the  most ge\-ne\-ral  spa\-tial\-ly 
homo\-ge\-neous and iso\-tropic one which, in terms of the 
\textit{comoving coordinates} 
$(t, r, \theta, \phi)$, has a line element reading as
\begin{equation}
\label{lfried}
ds^2=dt^2-R^2(t)\left(\frac{dr^2}{1-\kappa r^2}+
r^2d{\theta}^2+r^2\sin^2{\theta}d{\phi}^2\right)
\end{equation}
where the scale factor  $R(t)$ is a generic 
function of time only and, for an appropriate 
rescaling of the coordinates, the factor 
$\kappa=0,\pm1$  distinguishes the sign  
of constant spatial curvature. \\
%
%
 %
%
%
%
The FLRW dynamics is reduced to the time 
dependence of the scale factor $R(t)$, solving
 the Einstein equations 
with a diagonal stress-energy tensor $T_{\mu\nu}$ 
for all the fields present (matter, radiation, etc.); 
 for a 
perfect fluid it is characterized by a space-independent
energy density $\rho(t)$ and pressure $p(t)$ as 
given by
$	T^\mu{}_\nu=\textrm{diag}(\rho,-p,-p,-p)$;
in this case the $0-0$ component of the Einstein 
equations (the {\it Friedmann} one) and 
the $i-i$ components read as
\begin{align}
\label{frieq1}
\frac{{\dot R}^2}{R^2}+\frac{\kappa}{R^2}=
\frac{8\pi G}{3}\rho \, , \qquad
2\frac{\ddot R}{R}+\frac{{\dot R}^2}{R^2}+
\frac{\kappa}{R^2}=-8\pi Gp \, ,
\end{align}
respectively.
The difference between them leads to
	\begin{equation}
	\frac{\ddot R}{R}=-\frac{4\pi G}{3}(\rho+3p)	\, ,
	\end{equation}
which can be solved  for  $R(t)$ once provided an 
equation of state, i.e. a relation between  
$\rho$ and $p$. \\
When the Universe was radiation-dominated, 
as in the early period, 
the radiation component provided the greatest 
contribution to its energy density
and for a photon gas we have $p=\rho/3$. 
The present-time Universe  is, on the
contrary, matter-dominated: the ``particles'' 
(i.e. the galaxies) 
have only long-range (gravitational) interactions
and can be treated as a pressureless gas (``dust''):
the equation of state is $p=0$.

The energy tensor appearing in the 
Einstein field equations
describes the complete local energy due to all 
non-gravitational fields, while gravitational
energy has a non-local contribution. 
An unambiguous formulation for such a non-local 
expression is found only in the expressions
used at infinity for an asymptotically flat
space-time \cite{PEN82-ADM}. 
This is due to the property of the mass-energy 
term to be only one component of the 
energy-momentum tensor which can be reduced
only in a peculiar case
to a four-vector expression which can not 
be summed in a natural way. \\
Bearing in mind such difficulties,
the conservation law 
$T^{\mu\nu}{}_{;\nu}=0$  leads to
$	d(\rho R^3)=-pd(R^3)$, 
i.e. the first law of thermodynamics in an 
expanding Universe which, via the equation 
of state, leads to a differential relation for 
$\rho(R)$. 
A solution $\rho(R)$ for it 
in the Friedmann equation (\ref{frieq1}) 
corresponding to $\kappa=0$ provides the following 
behaviours
\begin{align}
\textrm{Radiation: \,} 
\rho \propto R^{-4}\, , \quad R \propto t^{1/2}  \,; \qquad 
\qquad	\textrm{Matter: \,} 
	\rho \propto R^{-3}\, , \quad R \propto t^{2/3}
\end{align}
where as long as the Universe is not
curvature-dominated, i.e. for sufficiently 
small values of $R$, 
the choice of $\kappa=0$ is not relevant.
Let us define the Hubble parameter $H\equiv {\dot R}/R$ and 
the critical density ${\rho}_c \equiv 3H^2/8\pi G$ in order
to rewrite (\ref{frieq1}) as
\begin{equation}
\label{fried3}
\frac{\kappa}{H^2R^2}=\frac{\rho}{3H^2/8\pi G}-1\equiv\Omega-1,
\end{equation}
where $\Omega$ is the ratio of the density 
to the critical one $\Omega\equiv\rho/{\rho}_c$;
since $H^2R^2$ is always positive, the relation 
between the sign of $\kappa$ and the sign of $(\Omega-1)$ 
reads
	\begin{align}
\textrm{Closed}:\,\,	\kappa=+1\, \Rightarrow \,\Omega>1 
												\quad ;\quad 
\textrm{Flat}: \,\,	\kappa=0\, \Rightarrow \, \Omega=1 
												\quad;\quad
\textrm{Open}:\,\,	\kappa=-1\, \Rightarrow \,\Omega<1  
	\end{align}


\subsection{Shortcomings of the SM: 
Horizon and Flatness Paradoxes\label{paradoxs}}

Despite the simplicity of the Friedmann solution 
 some paradoxes occur when 
taking into account the problem
of \textit{initial conditions}.
The observed Universe has to match very specific 
physical conditions in the very early epoch, but 
\cite{CH73} 
 showed that 
the set of initial data that can evolve to a 
Universe similar to the present one is of zero measure
and the standard model tells  {\it nothing} 
about initial conditions. \\
%

\textbf{Flatness.} \\
The value of $\Omega$ at present time $\Omega_0$
is related to the radius of curvature and to the density by 
Eq. (\ref{fried3}). The observational data restrict $\Omega_0$
to be of order of few units and consequently
$R_{curv}\sim H_0^{-1}$ and $\rho_0 \sim \rho_c$. 
By eliminating $H^2$ via the two equations in (\ref{fried3})
one finds the dependence of $\Omega$ on time as approaching unity for 
decreasing time as 
\begin{align}
\mid \Omega(10^{-43}~\textrm{ sec}) -1 \mid \lesssim \mathcal{O}(10^{-60})\, , \qquad
\mid \Omega(1 ~\textrm{sec}) -1 \mid \lesssim \mathcal{O}(10^{-16}) \, ,
\end{align}
and for the radius of curvature this would imply
\begin{align}
R_{\textrm{curv}}(10^{-43}~\textrm{ sec}) \mid \gtrsim 10^{30} H^{-1} \, , \qquad
R_{\textrm{curv}}(1 ~\textrm{ sec}) \mid \gtrsim 10^{8} H^{-1} \, .
\end{align}
These estimates imply that the FLRW model was characterized 
at the beginning by very special initial data in order to evolve in 
the Universe we observe today as, for example, 
a primeval deviation from the critical density 
when the temperature was  $T=10^{17}$ GeV (at the 
Grand Unification epoch) given by
\begin{equation}
	\left| \frac{\rho-\rho_c}{\rho} \right|_{T=10^{17}
	\mathrm{GeV}}<10^{-55} \, .
	\end{equation}
A flat Universe today requires $\Omega$ 
at ancient time close to 
unity up to a part in $10^{55}$. 
A little displacement from flatness at the beginning
-- for example $10^{-30}$ --
 would produce an actual Universe either very 
 open or very closed, so that $\Omega=1$ 
 is a very unstable condition: this is 
 the \textit{flatness problem}. \\
The natural time scale for cosmology is 
the Planck time
($\sim 10^{-43}$ sec): in a time of this order
a typical closed Universe would  reach 
the maximum size while an open one would become curvature 
dominated.
The actual Universe has survived $10^{60}$ 
Planck times without neither recollapsing nor becoming 
curvature dominated. \\

\textbf{Horizon.}\\
The other important problem arising from the SCM regards the 
explanation of the smoothness of the CMBR:  
the entire observable Universe could not have 
been in causal contact at remote time and eventually 
smoothed out, as a consequence of particle horizon.
In fact, a light signal (as a sample of the fastest
possible interaction) emitted at $t=0$ travelled 
during a  
time $t$ the physical distance 
	\begin{equation}
	\label{ldist}
	l(t)=R(t)\int_0^{\,t}{R(t')^{-1}}{dt'}=2\,t
	\end{equation}
in a radiation-dominated Universe with 
$R\propto t^{1/2}$, measuring  
the physical horizon size, i.e. the linear 
size of the greatest region causally connected 
at time $t$.
The  distance (\ref{ldist}) has to be compared with the 
radius $L(t)$ of the region which will evolve
in our currently observed part of the Universe. 
Conservation of entropy for $s\propto T^3$ 
gives
	\begin{equation}
	\label{ellel}
	L(t)=\left( {s_0}/{s(t)} \right) ^{1/3}L_0 \, ,
	\end{equation}
where $s_{0}$ is the present entropy density 
and $L_0\sim H^{-1}\simeq10^{10}$ years is 
the radius of the  observed Universe. 
An estimate of the ratio of the volumes provides,
as $T\sim 10^{17}\mathrm{GeV}$,  
	\begin{equation}
	\left. {l^3}/{L^3} \right|_{T=10^{17}\mathrm{GeV}} 
	\sim 10^{-83} \, .
	\end{equation}
The actual observable Universe is composed 
of several regions which have \textit{not} been in 
causal contact for the most part of their evolution,
preventing an explanation
about the present days Universe smoothness. 
In particular, the spectrum of the CMBR 
is uniform up to  $10^{-4}$. 
Moreover, we have at the time of 
recombination, i.e. when the photons of the CMBR last 
scattered, the ratio $l^3/L^3\sim 10^5$: 
the present Hubble volume consists of about 
$10^5$ causally disconnected
regions at recombination and no process could 
have smoothed out the temperature differences 
between these regions without violating causality. 
The particle horizon at recombination
subtends an angle of only $0.8^\circ$ in the sky 
today, while the CMBR is uniform across the sky.

\subsection{The Inflationary Paradigm}

The basic ideas  of the 
theory of inflation rely firstly on the 
original work \cite{G81}, 
i.e. the \textit{old inflation}, 
which provides a phase in the Universe evolution 
of exponential expansion; then  the
 formulation of \textit{new inflation} 
 by \cite{LA} 
 introduced the  
 slow-rolling phase in inflationary dynamics; 
finally,  many models  have sprung from 
the original theory (for a criticism, see \cite{AS04}.


In   \cite{G81} 
 is described a scenario  capable
of avoiding the horizon and flatness problems:
 both paradoxes would disappear 
dropping  the assumption of adiabaticity. 
In such a case, the entropy per comoving volume $s$
would be related as
$	s_0=Z^3 s_{\mathrm{early}}$, 
where $s_0$ and $s_{\mathrm{early}}$ refer
to the values at present and at very early times, 
for example at $T=T_0=10^{17}~\mathrm{GeV}$, 
and Z is some large factor. 
The estimate of the value
of $|\rho-\rho_c|/\rho$ is then multiplied by a factor $Z^2$ 
and would be of the  order of unity if 
$	Z>3\cdot 10^{27}$, 
getting rid of the flatness problem.\\
The right-hand side of (\ref{ellel}) is multiplied by
$Z^{-1}$: for  any
given temperature, the length scale 
of the early Universe is smaller by a factor $Z$ 
than previously evaluated, and for $Z$ 
sufficiently large the initial region which 
has evolved in our observed one would have 
been smaller than the horizon size at that time; 
 for $	Z>5\cdot 10^{27} $ 
the horizon problem disappears. \\
Making some \textit{ad hoc}
assumptions, the model accounts for the 
horizon and flatness paradoxes while a
suitable theory needs a physical process 
capable of such a large entropy production.
A simple solution relies on the assumption 
that at very early times  the energy
density of the Universe was dominated by 
a scalar field $\phi(\vec{x}, t)$,
i.e. 
$\rho=\rho_{\phi}+\rho_{\mathrm{rad}}+
\rho_{\mathrm{mat}}+\dots$ with
$\rho_\phi\gg\rho_{\mathrm{rad,\;mat,\;etc}}$
and hence $\rho\simeq\rho_{\phi}$. \\
The quantum field theory  Lagrangian density 
in this case and the corresponding stress-energy tensor read as 
	\begin{align}
	{\cal L}={\partial^{\mu}\phi\,
	\partial_{\mu}\phi}/2-V(\phi) \, , \quad 
	T^{\mu}_{~\nu} =\partial^{\mu}\phi\,
	\partial_{\nu}\phi-{\cal L}\delta^{\mu}_{~\nu} \, ,
	\end{align}
	resepctively, and hence for a perfect fluid
leads to 
\begin{align}	
\label{rphi}
	\rho_\phi={\dot{\phi}}/{2}^2+ V(\phi)+
	{R^{-2}}/2 \nabla^2\phi	\, , \qquad
	p_\phi={\dot{\phi}}/{2}^2-V(\phi)-
	{R^{-2}}/6\nabla^2\phi \, .
	\end{align}
Since spatial homogeneity implies a slow 
variation of $\phi$ with position, 
the spatial gradients are negligible and 
the ratio $\omega=p/\rho$ reads
\begin{equation}
\label{psur}
\frac{p_\phi}{\rho_\phi}\simeq	
         \frac{\displaystyle {\dot{\phi}}^2 /2 +V(\phi)}
{\displaystyle {\dot{\phi}}^2 /2 -V(\phi)} \, .
	\end{equation}
For a  field at a minimum of the potential $\dot{\phi}=0$
 and 
(\ref{psur}) becomes an equation of state as
$	p_\phi=-\rho_\phi$,
giving rise to a phase of exponential growth 
of $R\propto e^{Ht}$,  where the Hubble parameter 
$H$ remains constant: this is the {\it inflationary} 
or \textit{de Sitter} {\it phase}.\\
A very different evolution arises for a field  
in a thermal bath, in which case  the coupling can be summarized 
by adding a term $-(1/2)\lambda T^2\phi^2$ to 
the Lagrangian.   
The potential $V(\phi)$ is replaced by 
the \textit{finite-temperature} effective 
potential
	\begin{equation}
	\label{potphi}
	V_T(\phi)=V(\phi)+\lambda T^2\phi^2/2 \, .
	\end{equation}
This model can recorver the Standard Cosmology
via a phase transition of the scalar field between 
a metastable state (false minumum) and the true vacuum; 
the reheating due to oscillations around this 
state are damped by particle decay and, when the 
corresponding products thermalize, the Universe is reheated, 
and inflation comes to an end. This process 
(we will not discuss here the details)  
nevertheless leaves some problems open as, schematically:
%
%
(i) inflation never
			ends, due to smallness of  the tunnelling
			transition rate between the two minima; 
%
(ii) the phase transition is never completed;
%
(iii) the discontinuous process of bubble nucleation
				(exponential expansion of vacuum phases)
			via quantum tunnelling should produce a lot of 
			inhomogeneities which aren't actually observed.
%

\subsection{New Inflation: the Slow 
Rolling Model\label{slowrolling}}

In 1982, both \cite{LA} 
proposed a variant 
of Guth's model, now referred to
as \textit{new inflation} or 
\textit{slow-rolling inflation}, in order to
avoid the shortcomings
of the old inflation. 
Their original idea  considered 
a different mechanism of symmetry breaking, 
the so-called Coleman-Weinberg (CW) one, based 
on the gauge boson potential 
with a finite-temperature effective mass 
$	m_T\equiv\sqrt{-m^2+\lambda T^2} $ reading as
\begin{equation}
V_T(\phi)=\frac{B\sigma^4}2+
	B\phi^4\left[\ln\left(\frac{\phi^2}{\sigma^2}\right)
		-\frac12 \right]
			+\frac12 m^2_T\phi^2
	\label{CW} \, ,\end{equation}
where $B\simeq 10^{-3}$ is connected to the fundamental 
constants 
of the theory, while
$\sigma\simeq 2\cdot10^{15}$ GeV gives the energy 
associated to the 
symmetry breaking process. \\
The quantity ${m_T}^2$ can be used to 
parametrize the potential (\ref{CW}):
	\begin{enumerate}
	\item 	when ${m_T}^2>0$, the point $\phi=0$ 
			is a minimum of the 	potential, while when 
			$m_T^2<0$ it is a maximum;
	\item 	when ${m_T}^2 < 
	{4\sigma^2}/{e}\simeq 1.5\sigma^2$, 
		a second minimum develops for 
		some $\bar{\phi}>0$; initially 
		this minimum is higher than the one at 0, 
		but when $m_T$ becomes lower than
		a certain value $m_T^*$ $(0<m_T^*<1.5\sigma^2)$ 
		it will eventually become the global
	minimum of the potential.
	\end{enumerate}

If at some initial time the $\phi$-field is 
trapped in the minimum at $\phi=0$, 
the true minimum  
can  disappear
as the temperature lowers: 
 as $m_T$ approaches 0, the 
potential barrier becomes low and can 
 be  overcome by \textit{thermal} 
 (\textit{weakly} first order process)
 tunnelling, 
i.e. due to classical  fluctuations of 
the $\phi$ field around its minimum; 
the barrier can disappear completely when $m_T=0$
(second order process). 
The phase transition doesn't proceed via a
quantum tunnelling -- a very
discontinuous and a strongly first order 
process. 
The transition occurs rather smoothly,
avoiding the formation of undesired 
inhomogeneities. \\
When the $\phi$-field has passed the barrier
(if any), it begins to evolve towards its 
true minimum. 
The  model  (\ref{CW})
has the feature that if the 
coefficient of the logarithmic term is 
sufficiently 
high, the potential is very flat around 0 
and  the field $\phi$ ``slow rolls'' 
in the true vacuum state, rather than falling
abruptly: during this 
 phase the inflation takes place,
lasting enough to produce the required 
supercooling. When the field reaches the 
minimum, it begins to oscillate around it thus
originating the reheating. \\
The problems of Guth's originary model 
are skipped moving 
the inflationary phase \textit{after} the field
has escaped the false vacuum state, 
by adding the slow-rolling phase. \\
Virtually all models of 
inflation are based upon this principle.

\section{Evolution of Density Perturbations}

During the de Sitter phase the Hubble radius  
$H^{-1}$ is roughly constant while it 
increases during the FLRW phase. The scale factor
$R$ undergoes an $e$-folding in either case 
so that microphysics
(and then interaction between different close points)
can operate only at scale less then $\mathcal{O}(H^{-1})$. 
Hence, during the late inflationary epoch 
the evolution of the perturbations is essentially 
scale independent, since nothing can alter the 
amplitude of a real physical perturbation when its
scale is larger then $H^{-1}$.\\
The perturbed metric tensor  gives place to the 
equations \cite{BST83}

\begin{align}
\label{pp1}
[k^2 + 12 \pi G (\rho_0 +p_0)R^2]a = -k^2(1+3 c_s^2)h \, , 
\qquad 
(\dot{h} -Ha)^{\dot{}} + 3H (\dot{h} -Ha)=0 \, ,
\end{align}
where $h$ and $a$ denote two functions 
measuring the metric perturbations independent
of the choice of the spatial coordinates, 
$\rho_0$ and $p_0$ have to be considered
as the backgroungd energy density and pressure 
and $c_s^2 =dp_0/d\rho_0$ is formally the ``speed of sound'', 
while $k$ is the wave-number corresponding to the 
comoving wavelength $\lambda=2\pi/k$ (which is constant 
as the Universe expands). \\
The equations (\ref{pp1})
permit to identify, in the uniform Hubble constant 
gauge, the amplitude
\begin{align}
\label{zzeta}
\zeta \equiv h\left[1 + 
\frac{k^2}{12 \pi G (\rho_0+p_0)R^2} \right]
\end{align}
as a nearly time-independent quantity. 
Hence, by  the second of (\ref{pp1}),
there are two independent modes: a decaying one 
for which $\dot{h} -Ha \sim \textrm{exp}(-3Ht)$,
clearly negligible, and a constant one for which 
$\dot{h} = Ha$. This second mode, with a few 
calculations,  during inflation is expressed 
by the time-independence of the function (\ref{zzeta})
$\zeta= \zeta_f=const.$ and has the form  
$\zeta = {\delta \rho}/({\rho +p})$ \cite{GP82,BST83};
 its value has to 
match the final one $\zeta_f$ of the FLRW epoch.

\section{Inhomogeneous Inflationary Model}

In a synchronous reference, the generic line element
of a cosmological model takes the form
(in units $c=\hbar =1$)
\beq
ds^2 = dt^2 - \gamma_{\alpha\beta}(t, x^\mu)
dx^{\alpha }dx^{\beta } \, , \quad
\alpha,\beta,\mu= 1,2,3 
\label{a}
\eeq
where ${\gamma }_{\alpha \beta }(t, x^{\mu })$ 
is the three-dimensional metric tensor describing 
the geometry of the spatial slices. 
The Einstein equations
 in the presence 
of a self interacting scalar field 
$\{\phi (t, x^{\mu }), V(\phi)\}$ 
read explicitly  \cite{LK63}
\bseq
\label{einst}
\begin{align} 
& \frac{1}{2} \partial _t k_{\alpha }^{\alpha } + \frac{1}{4}
	k_{\alpha }^{\beta }k_{\beta }^{\alpha } = 
	\chi \left[ - (\partial _t\phi )^2 + V(\phi )\right] 
	\label{b}  \\
& \frac{1}{2}(k^{\beta }_{\alpha ;\beta } - 
	k^{\beta }_{\beta ;\alpha }) = 
	\chi \left(\partial _{\alpha }\phi \,\partial _t \phi \right) 
	\label{c} \\
& \!\!\!\!
	\frac{1}{2\sqrt{\gamma }}\partial _t (\sqrt{\gamma }\,
	k_{\alpha }^{\beta }) + P_{\alpha }^{\beta } = 
	\chi \left[ {\gamma }^{\beta \mu } \,
	\partial _{\alpha }\phi \,\partial _{\mu }\phi + 
 	V(\phi ) \,{\delta }_{\alpha }^{\beta }
	\right] \, ,
	\label{d} 
\end{align}
\eseq
where $\chi = 8\pi G$, the three-dimensional 
Ricci tensor $P_{\alpha }^{\beta }$
is constructed via 
${\gamma }_{\alpha \beta}$, 
$\gamma \equiv {\mathrm det} ~\gamma_{\alpha \beta} $,
$k_{\alpha \beta } \equiv \partial _t{\gamma }_{\alpha \beta } $.
The dynamics of the scalar field 
$\phi (t, x^{\gamma })$ is coupled to the system (\ref{einst})
and is described by the equation 
\beq 
\partial _{tt}\phi + \frac{1}{2}k_{\alpha }^{\alpha }
\partial _t\phi - {\gamma }^{\alpha \beta }{\phi }_{;\alpha ;\beta } 
+ \frac{dV}{d\phi } = 0 \, .
\label{m} 
\eeq
In what follows we will consider the  
three fundamental statements:
\begin{enumerate}
\renewcommand{\labelenumi}{({\it\roman{enumi}} )}
	\item \label{prima}
	 the three metric tensor is taken in the general
	 factorized form 
	 $	 {\gamma }_{\alpha \beta }(t, x^{\mu })=
	 \Gamma^2(t, x^{\mu })\xi_{\alpha \beta }
	 (x^{\mu })$,
where $\xi_{\alpha \beta }$ is a generic 
symmetric three-tensor and therefore 
contains six arbitrary functions of the spatial 
coordinates, while $\Gamma$ is to be determined 
by the dynamics;

	\item \label{seconda}
	the self interacting scalar field
	dynamics is described by a potential term 
	which satisfies all the features of an inflationary
	one, i.e. a symmetry breaking confi\-gu\-ration
	characterized by a relevant plateau region; 
	\item \label{terza}
	the inflationary solution is constructed
	under the assumptions
	\bseq
	\label{infla12}
	\begin{align}
  \label{infla1}
  &{1}/{2}\left(\partial_t \phi\right)^2 \ll 
  V\left(\phi\right) \\
 & \mid \partial_{tt}\phi \mid \ll \mid 
  k^{\alpha}_{\alpha} \, \partial_t\phi \mid
  \label{infla2} \, .
  \end{align}
  \eseq
\end{enumerate}

Our analysis, following \cite{IMmpla04},
 concerns the evolution of the 
cosmological model when the scalar field slow 
rolls on the plateau and the corresponding 
potential term is described as
	\beq
	\label{pot}
	V(\phi) = \Lambda_0-\lambda U(\phi) \, ,
	\eeq
where $\Lambda_0$ behaves as an effective 
cosmological constant  of the order 
$10^{15}-10^{16}~\mathrm{GeV}$ and $\lambda$
($\ll 1$) is a coupling constant associated
to the perturbation $U(\phi)$.\\
Since the scalar field moves on a 
plateau almost flat, we infer that to lowest 
order of approximation 
$\phi(t,x^{\gamma})\sim \alpha(x^{\gamma})$ 
(see below (\ref{fi})) and
therefore the potential reduces to 
a space-dependent effective cosmological 
constant 
\beq
\label{pot2}
\Lambda(x^{\gamma})\equiv \Lambda_0
-\lambda U(\alpha(x^{\gamma})) \, .
\eeq
In this scheme the $0-0$ (\ref{b}) and 
$\alpha-\beta$ (\ref{d})
components of the Einstein equations 
reduce, under condition
{\textit{(iii)} and neglecting all the spatial 
gradients, to the simple ones
\begin{align}
3\, \partial_{tt}\ln \Gamma+3\, 
(\partial_t \ln \Gamma)^2= 
\chi \Lambda(x^{\gamma}) \, ,\qquad 
(\partial_{tt}\ln \Gamma )\delta^{\alpha}_{\beta} +
3\, (\partial_t \ln \Gamma)^2\delta^{\alpha}_{\beta} =
\chi \Lambda(x^{\gamma})\delta^{\alpha}_{\beta} \, ,
\label{00ab}
\end{align}
respectively.
A simultaneous solution for $\Gamma$ of both 
equations (\ref{00ab}) takes the form
\beq
\label{ve}
\Gamma(x^{\gamma})=\Gamma_0(x^{\gamma})
\exp\left[
\sqrt{
{\chi\Lambda(x^{\gamma})}/{3}}(t-t_0)\right] \, ,
\eeq
where $\Gamma_0(x^{\gamma})$ is an integration function
while $t_0$ a given initial instant of time for the 
inflationary scenario.
Under the same assumptions and taking into account 
(\ref{ve}) for $\Gamma$, the scalar field 
equation (\ref{m}) rewrites as 
\begin{align}
\label{vd}
3H(x^{\gamma}) \partial_t \phi - \lambda W(\phi)=0 \, ,
 \qquad\qquad
H(x^{\gamma})=\partial_t \ln \Gamma
=\sqrt{{\chi}\Lambda(x^{\gamma})/{3}} \, , \quad 
 W(\phi)={dU}/{d\phi} \, . 
\end{align}
We search for a solution of the dynamical equation
(\ref{vd}) in the form
\beq
\label{fi}
\phi(t,x^{\gamma})=\alpha(x^{\gamma})+
\beta(x^{\gamma})\left( t-t_0\right) \, .
\eeq
Inserting expression (\ref{fi}) in (\ref{vd})
and considering it to the lowest order,
it is possible  to express 
$\beta$ in terms of $\alpha$ as
\beq
\label{beta}
\beta=\frac{\lambda W(\alpha)}{\sqrt{3\chi 
\Lambda_0 - \lambda U(\alpha)}} \,.
\eeq
Of course the validity of solution (\ref{beta})
takes place in the limit
\beq
\label{lim}
t-t_0 \ll \left| \frac{\alpha}{\beta}
\right| = \left|\frac{\alpha}{ W(\alpha)}
\sqrt{3\chi \frac{\Lambda_0}{\lambda^2} -
\frac{U(\alpha)}{ \lambda}}\,\right|
\eeq
where the ratio $\Lambda_0/\lambda^2$ takes in 
general very large values.\\
The $0-\alpha$ component (\ref{c}), in view of (\ref{ve})
and (\ref{fi}) through (\ref{beta}),
reduces to
$\partial_{\gamma}
\left( \Lambda + \lambda U\right) =0 $,
which for $\Lambda(x^{\gamma})$ is an identity 
as (\ref{pot2}). \\
The 
spatial gradients, either of the three-metric
field either of the scalar one, behave 
as $\Gamma^{-2}$ and  decay 
exponentially. 

If we take into account the coordinate
characteristic lengths $L$ and $l$
for the inhomogeneity scales 
regarding the functions
$\Gamma_0$ and $\xi_{\alpha \beta}$, 
i.e.
\beq
\partial_{\gamma} \Gamma_0  
\sim {\Gamma_0}/{L} \, , \qquad
\partial_{\gamma} \,\xi_{\alpha \beta} 
\sim {\xi_{\alpha \beta}}/{l}\, ,
\eeq
respectively, negligibility of the spatial 
gradients
at the beginning leads to the 
inequalities for the phy\-sical quantities
\begin{align}
\Gamma_0 l = l_{\textnormal{phys}} \gg H^{-1} \, , \qquad
\Gamma_0 L = L_{\textnormal{phys}} \gg H^{-1} \, .
\label{h-1}
\end{align}
These conditions state that all the inhomogeneities
have to be much greater then the physical 
horizon $H^{-1}$. \\
Negligibility of the 
spatial gradients at the beginning of 
inflation is required (as well known)
by the existence of the de Sitter phase itself;
however, spatial gradients having a passive 
dynamical role allow to deal with a 
fully  inhomogeneous solution: space point 
dynamically decouple to leading order.

The condition (\ref{infla1}) is naturally 
satisfied since 
states that the dominant contribution 
in $\Lambda(x^{\gamma})$ is provided by 
$\Lambda_0$, i.e. $\lambda U(\alpha) \ll \Lambda_0$;
the same can be said for (\ref{infla2}). \\
The only important restriction 
on the spatial function $\alpha(x^{\gamma})$
is 
$|\alpha| \ll \left|U^{-1}
\left({\Lambda_0}/{\lambda}\right)\right|$.\\
A satisfactory 
exponential expansion able to overcome 
the SCM shortcomings, i.e. to freeze the density 
fluctuations between the beginning and the end 
of the inflation, requires
that in each space point the condition for the $e$-folding
$H(t_f - t_i) \sim \mathcal{O}(10^2) $
holds, where $t_i$ and $t_f$ denote 
the instants when the de Sitter phase starts 
and ends, respectively. 
We may take $t_i\equiv t_0$ 
and hence for 
$t_f$  we have, by (\ref{vd}) and (\ref{beta}),
\beq
\label{ineq2}
H (t_f - t_i) \ll ({\Lambda_0}/{\lambda}) ~
\left|{\alpha}/{W(\alpha)}\right| \, .
\eeq 
Since $\Lambda_0/\lambda$ is a very large quantity, 
no serious restrictions appear for the e-folding 
of the model.

The obtained solution possesses a very general festure: in fact, 
once sa\-tis\-fied all the dynamical equations, still 
eight arbitrary spatial functions remain, i.e.
six for $\xi_{\alpha \beta}(x^{\gamma})$, and then
$\Gamma_0(x^{\gamma})$, $\alpha(x^{\gamma})$.
However, taking into account the possibility 
to choose an arbitrary gauge via the set 
of the spatial coordinates, we have to kill 
three degrees of freedom; hence  
five physically arbitrary functions finally remain:
four corresponding to gravity degrees of freedom
and one related to the scalar field.\\
This picture corresponds exactly to the allowance
of specifying a generic Cauchy problem for the 
gravitational field, on a spatial non-singular
hypersurface, nevertheless one degree of freedom 
of the scalar field is lost against the full 
generality.

\section{Applications}

\subsection{Coleman--Weinberg Model}

Let us specify this solution in the case of the
 vanishing mass zero-temperature CW
potential
\cite{CW73}
\beq
\label{cw}
V(\phi)=\frac{B\sigma^4}{2}
+B\phi^4\left[\ln\left(\frac{\phi^2}{\sigma^2}\right)
-\frac12 \right] \, .
	\end{equation}
In the region $\mid \phi\mid\ll \mid \sigma\mid$
the potential (\ref{cw}) approaches a plateau 
 profile similar to (\ref{pot}) 
and acquires the form
\beq
\label{cwa}
V(\phi)\simeq {B\sigma^4}/{2} 
-{\lambda}\phi^4/{4} \, , 
\qquad \lambda\simeq80 B\simeq 0.1 \, .
\eeq
This is  reducible to (\ref{pot}) by 
setting 
$ \Lambda_0= {B\sigma^4}/{2}$, 
$U(\phi)={\phi^4}/{4}$ and 
$ W(\phi)=\phi^3 $. 
Hence the relations (\ref{beta}) 
rewrites as $\beta={\lambda \alpha^3}/{3H}$
while 
the inequality for $\alpha$ is equivalent
to fulfil the initial assumption
$\Lambda_0 \gg 
\lambda U(\alpha)\sim {\lambda}\alpha^4/{4} $,
that is 
\begin{align}
\label{asig}
|\alpha| \ll 
\sqrt[4]{{\Lambda_0}/{\lambda}} \sim \sigma \, .
\end{align}

\subsection{Lemaitre--Tolman Spherically Symmetric Metric}

It is interesting to consider the framework 
developed so far when in presence of a spherically 
symmetric line element written as
\beq
ds^2= dt^2- e^{2 \pi} dr^2 - e^{2 \psi}d\Omega
\eeq
where $\pi$ and $\psi$ are functions of 
$t$ and $r$ only. 
The corresponding Einstein equations  read as
\bseq
\label{Teinst}
\begin{align}
&(\dot{\psi})^2 + 2 \dot{\pi}\dot{\psi}+
e^{-2 \psi} - e^{-2 \pi}\left[  
2 \psi^{\prime \prime} + 3 (\psi^{\prime})^2
- 2 \pi^\prime \psi^\prime \right] =
\frac{\chi}{2}  \left[  (\partial _t\phi )^2  
+(\partial _r\phi )^2 e^{-2\pi} + V(\phi )\right] 
\label{Tb}  \\
&2 \ddot{\psi } +3 (\dot{\psi})^2 
+ e^{-2 \psi} -(\psi^{\prime})^2e^{-2\pi}
 = -\frac{\chi}{2} \left[  (\partial _t\phi )^2  
+(\partial _r\phi )^2 e^{-2\pi} - V(\phi )\right]
\label{Tc} \\
&2 {\dot{\psi}}^{\prime} + 2 \psi^{\prime}\dot{\psi}
-2 \dot{\pi}\psi^{\prime}=
\chi \dot{\phi} \phi^{\prime} \, , 
\label{Td} 
\end{align}
\eseq
while the scalar field 
$\phi (t, x^{\gamma })$ 
is described by the equation 
\beq 
\partial _{tt}\phi + 
\left[ \dot\pi +2\dot\psi\right]\partial _t\phi 
- \left[ \phi^{\prime \prime}  
+\phi^{\prime}\left(-\pi^\prime + 2 \psi^\prime \right)
\right] e^{-2\pi}
+ \frac{dV}{d\phi } = 0 \, .
\label{Tm} 
\eeq
Once defined $e^\psi\equiv r m(r,t) $,
without entering in the details of the 
derivation \cite{Itolman04}, 
we can find again for large values of $m$
\beq
\sqrt{\frac{2\chi \Lambda}{3}}m=
e^{ \sqrt{{\chi \Lambda}/{6}} ~
(t-t_0) } \, ,
\eeq
which admits linear solution for the 
scalar field similar to (\ref{fi}),
satisfying  the set of Einstein equations
under the same assumptions of the previous discussion.

\subsection{Towards FLRW Universe}

The conditions (\ref{h-1}) state that the validity 
of the inhomogeneous inflationary scenario discussed 
so far requires the 
inhomogeneous scales to be out of the horizon 
when inflation starts. 
The situation is different when treating
the small perturbations to the FLRW case; 
in fact, the negligibility of the spatial curvature 
corresponds to require the radius of curvature 
of the Universe to be much greater than the 
physical horizon,  the inhomogeneous terms 
being small in amplitude. 
To this end, let us consider the 
three-metric 
\beq
\gamma_{\alpha \beta} = \Gamma(t,\varphi^{\mu})^2 
\left[h_{\alpha \beta} + (t-t_0)
\delta	\theta_{\alpha \beta}(\varphi^{\mu}) 
	\right] \, , 
	\label{tfw}
\eeq
where $h_{\alpha\beta}$	denotes the FLRW spatial 
part of the three metric 
($\{\varphi^{\mu}\}$
are the three usual angular coordinates) 
and  $\delta	\theta_{\alpha \beta}$ denote 
a small inhomogeneous perturbation. 
The Einstein equations (\ref{einst})
coupled to the scalar field dynamics  (\ref{m})
on the plateau (\ref{cwa}) admit,  
to leading order in the inhomogeneities,
 the solution
\bseq
\begin{align}
&\Gamma=\Gamma_0 e^{H(t-t_0)} \, , \qquad\qquad \qquad 
H= H_0 - \frac{\delta	\theta}{6} \, , \qquad
 H_0 = \sigma^2 \sqrt{\frac{\chi B}{6}}  \, ,\\
&
\phi= \alpha_0 \left[ 1+ 
	\frac{\lambda \alpha_0^2}{3H_0}(t-t_0) \right]
	+ \frac{\delta	\theta}{3 \chi}
  \left[ 1+ 
	\frac{\lambda \alpha_0^2}{H_0}(t-t_0) \right] \, ,\qquad
 \delta	\theta_{\alpha \beta} = 
 	\frac{\delta\theta}{3} h_{\alpha \beta} \, ,
\end{align}
\label{solufrw}
\eseq
where $t_0$ and $ \Gamma_0$ are constants.
The solution (\ref{solufrw}) holds and 
provides the correct $e$-folding of order 
$\mathcal{O}(10^2)$ when the  
inequalities 
\bseq
\begin{align}
& t-t_0 \ll {3 H_0}/{\lambda \alpha_0^2} \, , 
\label{di1} \, \qquad \quad
 \alpha_0 \ll \mathcal{O} 
	\left( {10^{-1}}
			\sqrt{{\Lambda_0}/{\lambda}}\right) \, ,
			\\
\end{align}
take place, the first one  ensuring that the 
dominant term of the scalar field remains the 
time-independent one during the de Sitter phase, while 
the second one  
accounts for the $e$-folding;
the   spatial gradients 
in the Einstein  field equations 
behave as  (see, for instance, the RHS 
of Eq. (\ref{d}))
$\Gamma^{-2}\partial_\mu\alpha\partial_\nu\alpha \sim
\Gamma^{-2}\delta^2/l^2$ which have to be compared
to the potential 
term $V\sim \lambda \alpha^4$ and therefore
are negligible, while 
the  case for the scalar field  is similar. 
Hence we get
 
\begin{align}
&\label{di3} 
 \Gamma_0 l = l_{\textrm{phys}} \gg 	H_0^{-1}\delta \, , \qquad 
	l_{\textrm{phys}} \gg {\delta}/{\sqrt{\lambda \alpha_0^3}}\, ,
\end{align}
\label{diseq}
\eseq
respectively,
where $\delta$ 
$(\ll  H_0/100)$
and $l$ in (\ref{di3}) denote the 
cha\-racte\-ristic amplitude and length, 
 of the \textit{arbitrary} function
$\delta \theta$, trace of the 
tensor $\delta \theta_{\alpha \beta}$. 
When inflation starts the inhomogeneous 
scales can be inside the physical horizon 
$H_0^{-1}$.

The physical implications on the 
density perturbation spectrum of such a nearly 
homogeneous model  rely on the 
dominant behaviour of the potential term 
over the energy 
density $\rho_{\phi}$ associated to the 
scalar field during the de Sitter phase  
and therefore
\beq
\label{drho}
\Delta\equiv 
\left|\frac{\delta\rho_{\phi}}{\rho_{\phi}}\right|	
\sim \left|\frac{d\ln V}{d\phi}\delta\phi \right|
\simeq \left|\frac{\lambda}{\Lambda_0}\,
W(\alpha_0)\,\delta\alpha \right| \, ,
\eeq	 
where $\delta\alpha= \delta\theta/(3\chi)$ 
for our scalar field solution (\ref{solufrw}).
In particular, in the CW
case, Equation (\ref{drho}) reduces to 
\beq
\label{cwa1}
\Delta_{CW}\simeq \frac{50}{\sigma^4}
{\alpha_0}^3 \, \frac{\delta\theta}{\chi} \, .
\eeq
However, to compute the physically 
relevant perturbations after the scales re-entry 
in the horizon, let us evaluate  
the gauge invariant quantity $\zeta$ 
which has now the form (\ref{zzeta})
\beq
\label{zeta}
\zeta = \frac{\delta \rho}{\rho +p}\cong 
\frac{\delta \theta}{W(\alpha_0)}
\frac{\Lambda_0}{\lambda} 
\eeq
when the perturbations 
leave the horizon and
$\rho + p = (\partial_t \phi)^2$ (see Eq. \ref{rphi});
in the CW case it reads as 
\beq
\label{zetacw}
\zeta_{CW}=  \frac{\sigma^4}{160}~
\frac{\delta \theta}{{\alpha_0}^3} \, .
\eeq
Since $\zeta$ remains constant during the 
super-horizon evolution of the perturbations, 
then at the re-entry to the causal scale in the 
matter-dominated era we get
$\zeta_{MD} \sim \delta \rho/\rho \sim \zeta_{CW}$.\\
By restoring physical units and assuming 
$\alpha_0 \lesssim  10^{-4}\sigma/\sqrt{hc}$ 
in agreement with (\ref{asig}), then 
 it is required  
$\delta \alpha/\alpha_0 \lesssim 10^{-2}$
in order to obtain perturbations 
$\delta \rho/\rho \sim 10^{-4}$ 
at the  horizon re-entry  
during  the matter-dominated age. \\
Hence the expression (\ref{zetacw}) explains how
the perturbation 
spectrum after the de Sitter phase can 
still arise from classical inhomogeneous 
terms. Indeed, the function 
$\delta\theta(\varphi^{\mu})$ is an arbitrary one
and can be chosen for it a Harrison--Zeldovich
spectrum by assigning its Fourier transform as 
\beq
\label{hz}
\left|\delta \alpha(k)\right|^2\propto 
{\textnormal{const.}}/{k^{3}} \, ;
\eeq
such a spectrum has to  hold for 
$k\ll {\Gamma_0}/{( H_0^{-1}\delta)}$.

Thus, the pre-inflationary 
inhomogeneities of the scalar field
remain almost of the same amplitude 
during the de Sitter phase as a consequence
of the linear form of 
the scalar field solution (\ref{fi}). 
We get that the Harrison--Zeldovich 
spectrum can be a 
pre-inflationary picture of the density perturbations
and it survives to the de Sitter phase, becoming
a classical seed for structure formation.
The existence of such a classical spectrum is not
related with the quantum fluctuations 
of the scalar field, whose 
effect is an independent contribution to the 
classical one.

\section{Concluding Remarks}

The merit of our analysis relies on 
having provided a dynamical framework 
within which classical inhomogeneous 
perturbations to a real scalar field 
minimally coupled with gravity can 
survive even after that the de Sitter expansion 
of the Universe stretched the geometry;
the key feature underlying this result
consists \textit{(i)} of constructing an 
inhomogeneous model for which the leading
order of the scalar field is provided 
by a spatial function and then  \textit{(ii)}
of showing how the very general case contains 
as a limit a model close to the FLRW one. 
Moreover such results have been extended
to the spherically symmetric Lemaitre--Tolman
metric.

It is relevant to remark that the metric 
tensor (\ref{tfw}) seems of the same form 
as the one considered in \cite{IM03}; 
however in the present paper the function 
$\eta(t)$ appearing in the previous work
is linear in time and does not decay 
exponentially. The different behaviour
relies on the negligibility of the matter 
with respect to the scalar field which 
is at the ground of the present analysis. 
We are here assuming the dynamics of $\eta(t)$
to be driven by the scalar field alone,
instead of by the ultra-relativistic matter.
This situation corresponds to an initial 
conditions for which the scalar field dominates
over the ultra-relativistic matter when 
inflation starts and this is the reason 
for the resulting different issues of the two 
analyses.





\section*{Acknowledgements}

This work has been done 
when visiting the Maths Department at Queen 
Mary, University of London, benefiting of the 
kind hospitality of the General Relativity 
and Cosmology group, under grant funded partially  by 
``Accademia Nazionale dei Lincei - Royal Society''
and partially by Fondazione ``Angelo Della Riccia''. \\
%



\begin{thebibliography}{99}
%
%


\bibitem{KT90} 
E.W. Kolb and M.S. Turner, 
\textit{The Early Universe},
(Adison-Wesley, Reading) (1990).


\bibitem{LK63} 
E.M.Lifshitz and I.M.Khalatnikov, 
\textit{Adv.Phys.}  
\textbf{12}, 185 (1963).

 
\bibitem{dB01-map} 
P. de Bernardis et al., 
{\it Astrophys.J.}  {\bf 564}, 559, (2002);
%
 %
D.N. Spergel, 
\textit{Astrophys.J. Suppl. Series} (2003)
{\bf 148}, 175
 
\bibitem{G81}  A.H. Guth, {\it Phys. Rev. D},  
\textbf{23}, 347 (1981).

\bibitem{infla}
 A.D. Linde, \textit{Phys. Lett.}, 
\textbf{108B}, 389 (1982);
%
A.A. Starobinsky, {\it JETP Lett.}, 
{\bf 30}, 682 (1979);
%
A.A. Starobinsky, {\it Phys. Lett. B},
{\bf 91}, 99 (1980);
%
A.D. Linde, \textit{Phys. Lett.}, 
\textbf{129B}, 177 (1983);
%
J. Silk and M.S. Turner, 
\textit{Phys. Rev. D},  
\textbf{35}, 419 (1987).

\bibitem{GP82} A.H. Guth and S. Pi \textit{Phys. Rev. Lett}
{\bf 49}, n.15 1110 (1982)

\bibitem{BST83} J.M. Bardeen, P.J. Steinhardt and 
M.S.Turner \textit{Phys.Rev.D} \textbf{28}, n.4, 679 (1983)





\bibitem{MB95} C.P.Ma and E. Bertschinger, 
\textit{Astrophys.J.},
 \textbf{455}, 7 (1995).

\bibitem{DT94} K. Tomita and N. Deruelle, 
\textit{Phys.Rev.D}, 
 {\bf 50}, 7216 (1994).


\bibitem{IM03}
G. Imponente and G. Montani
 {\it Int. Journ. Mod. Phys. D}, 
 \textbf{12}, n.10, 1845
(available gr-qc/0307048) (2003).

\bibitem{STA96} D. Polarski and A.A. Starobinsky, 
{\it Class. Quant. Grav.}, {\bf 13}, 377 (1996).

\bibitem{IMmpla04}
G. Imponente and G. Montani
\textit{Mod. Phys. Lett. A}, \textbf{19}, n.17, 
1281-1290 (2004) (available gr-qc/0404031).

\bibitem{MAC79-BKL82} 
M.A.H. MacCallum, 
In ``General relativity: an Einstein centenary survey'', 
ed. S.W. Hawking and W. Israel, p.533. Cambridge
University Press (1979);
%
V.A. Belinski, I.M. Khalatnikov and E.M. Lifshitz, 
\textit{Adv. Phys.}  \textbf{31},  639 (1982).

\bibitem{SKB}
A.A. Starobinsky, 
\textit{Pis'ma Zh.Eksp.Teor.Fiz.} 
\textbf{37}, 55 (1983);
%
%
A.Kirillov and G. Montani, 
{\it Phys.Rev.D} {\bf 66}, 064010 (2002);
%
%
J.D. Barrow, 
\textit{Phys. Lett. B} \textbf{187}, 12 (1987).


\bibitem{LA}
A.D. Linde,
{\it Phys. Lett.}, {\bf 108B}, (1982) 389;
%
A. Albrecht and P. J. Steinhardt,
{\it Phys. Rev. Lett.}, {\bf 48}, (1982) 1220.

\bibitem{AS04}
A. Albrecht and L. Sorbo, 
to appear in \textit{Phys. Rev. D} (2004), 
(available  hep-th/0405270), 

\bibitem{PEN82-ADM}
R. Penrose (1982), \textit{Proc.R.Soc.Lond.}
\textbf{A 381}, 53;
%
R. Arnowitt, S. Deser and C.W. Misner, in {\it Gravitation: 
an introduction to current research} (1962), eds. 
I. Witten and J. Wiley, New York.

\bibitem{CH73} C.B. Collins and S.W. Hawking, (1973)
\textit{Astrophys. Journ.} \textbf{180}, 317


\bibitem{M00}  G. Montani, 
\textit{Class. and Quantum Grav.}  
\textbf{17}, 2205 (2000).


\bibitem{CW73}  S.Coleman and E. Weinberg, 
\textit{Phys.Rev.D} \textbf{7}, 1888 (1973).



\bibitem{Itolman04} G. Imponente, in preparation




\end{thebibliography}
\end{document}